\documentclass[a4paper,USenglish,cleveref, autoref, runningheads]{article}

\usepackage{todonotes}
\usepackage{xspace}
\usepackage{bm, bbm}
\usepackage[shortlabels]{enumitem}
\usepackage{algorithm}
\usepackage[noend]{algorithmic}
\usepackage{wrapfig}
\usepackage{color}
\usepackage{colortbl}
\usepackage{amsmath}
\usepackage{amsfonts}
\usepackage{subcaption}
\usepackage{url}

\newcommand{\bsb}{\textsc{BSB}\xspace}
\newcommand{\halfsplit}{\textsc{Half-split}\xspace}
\newcommand{\maxdemand}{\textsc{Max-demand}\xspace}
\DeclareMathOperator{\dis}{dist}
\newcommand{\quoteMe}[1]{``#1''}

\title{BSB: Towards Demand-Aware Peer Selection With XOR-based Routing} 

\author{Qingyun Ji \and
Darya Melnyk \and
Arash Pourdamghani
\and Stefan Schmid \\ TU Berlin, Germany}

\date{}

\begin{document}

\maketitle

\begin{abstract}
Peer-to-peer networks, as a key enabler of modern networked and distributed systems, rely on peer-selection algorithms to optimize their scalability and performance. Peer-selection methods have been studied extensively in various aspects, including routing mechanisms and communication overhead. However, many state-of-the-art algorithms are oblivious to application-specific data traffic. This mismatch between design and demand results in underutilized connections, which inevitably leads to longer paths and increased latency.

In this work, we propose a novel demand-aware peer-selection algorithm, called \textit{Binary Search in Buckets} (\bsb). Our demand-aware approach adheres to a local and greedy XOR-based routing mechanism, ensuring compatibility with existing protocols and mechanisms. We evaluate our solution against two prior algorithms by conducting simulations on real-world and synthetic communication network traces. The results of our evaluations show that \bsb can offer up to a $43\%$ improvement compared to two selected algorithms from the literature. 
\end{abstract}


\section{Introduction}
Many modern networked and distributed systems rely on peer-selection algorithms, including (but not limited to) Amazon Dynamo~\cite{Dynamo07}, Apache Cassandra~\cite{Cassandra10}, and Ethereum~\cite{KifferSLMN21}. 
Since peer-selection affects the latency and throughput of the system, designing efficient and simple peer-selection algorithms has been a major research topic over the past decades~\cite{continuous-discrete03}. 
While many different algorithms have been proposed in the literature already, they are typically oblivious to the inherent patterns in demand.  
Recently, Avin et al.~\cite{AvinGGS20} have demonstrated that different types of applications have their own unique data traffic characteristics. These characteristics manifest, for example, in significant spatial (non-temporal) locality. Roughly speaking, high spatial locality means that communication happens frequently between specific pairs of nodes within the network. 
This work takes a step toward demand-aware peer-selection, by introducing a novel approach that utilizes information about the communication demand to optimize peer selection, while also utilizing XOR-based routing~\cite{MaymounkovM02}. 
This change from demand-oblivious to demand-aware design promises a more efficient, scalable, and cost-effective network topology.

\subsection{Our contribution}
In this paper, we present two novel demand-aware peer-selection algorithms, under the class of \textit{Binary Search in Buckets} (\bsb). These algorithms use the fact that communication between peers reveals patterns that can be exploited by the peer-selection process. The algorithms are based on the structure of Kademlia~\cite{MaymounkovM02}. From the perspective of a sending peer, the peers of the network are divided into buckets, where the number of peers in each bucket increases exponentially. The algorithms differ in how sending peers select which peers to communicate with inside the buckets.
\begin{itemize}
    \item The \halfsplit strategy considers the demand in each bucket and connects to a central peer based on the demand.
    \item The \maxdemand strategy connects to the peer with the maximum demand in each bucket. 
\end{itemize}
By design, our algorithms respect the local and  greedy XOR-based routing and rely on a local view of the demand. Hence, they can replace most of the existing peer-selection algorithms without requiring a change in other parts of the system.
We show the benefits of our algorithms by comparing them to previous algorithms on a range of synthetic and real-world datasets. We show that for the cases where the spatial complexity of demand is not high, our algorithms can perform better than the previous algorithms, which is observed in some real-world instances. 

\subsection{Related work}
In this section, we first review the traditional algorithms for peer selection, and then discuss works related to demand-aware designs

\noindent \textbf{Peer-selection algorithms.} One of the key results in peer-to-peer (P2P) network design is Chord~\cite{StoicaMLKKDB03}, a simple network design for efficient network location. The peer identifiers are arranged on a ring, each peer maintains up to $ \log n$ connections, and key location is implemented via greedy routing. The Kademlia protocol~\cite{MaymounkovM02} improves the key location by allowing the peers to select their connections in a randomized manner. 
Many more demand-oblivious P2P network designs have been proposed in the literature, including Viceroy~\cite{MalkhiNR02},  
CAN~\cite{RatnasamyFHKS01}, Pastry~\cite{RowstronD01}, and Tapestry~\cite{ZhaoHSRJK04}.
For a detailed overview of P2P network architectures, please refer to~\cite{Androutsellis-TheotokisS04} and~\cite{LuaCPSL05}.

\noindent \textbf{Demand-aware network design.} 
Demand-aware network design has been pioneered by the work of~\cite{AvinS18}, which gives a glimpse of the potential and challenges of demand-aware network design. Following this work, demand-aware network design has been studied in the context of datacenter networks~\cite{HanauerHST22,PourdamghaniASS23,FigielMM025}, distributed hash tables~\cite{PourdamghaniASS24}, and other settings~\cite{AddankiPPRSV23,ifip25}.  
One direction closely related to our work is presented in~\cite{AvinMS20,FigielKO024,LASLiN}. There, the authors consider the problem of designing \emph{bounded-degree} demand-aware networks, i.e., demand-aware networks where each peer can only be connected to a constant number of other peers. 
More recently, authors of~\cite{alenex25} provided algorithms for demand-aware augmentation of an existing network with a matching. 

\begin{table}[tb]
	\centering
		\begin{tabular}{|c|c|c|c|}
			\hline
			\textbf{Protocol} & \textbf{Demand-awareness} & \textbf{XOR-based routing} \\
			\hline  
            Chord~\cite{StoicaMLKKDB03} & \cellcolor{red!20} No & \cellcolor{green!20} Yes \\ \hline 
            Permutations~\cite{WangKZGJM0K23} & \cellcolor{green!20} Yes &  \cellcolor{red!20} No
            \\ \hline
            \bsb [this paper] & \cellcolor{green!20} Yes & \cellcolor{green!20} Yes
            \\ \hline 
			
		\end{tabular}
	\caption{A summary of selected peer-selection algorithms compared to our work.}
	\label{table:DHT}
\end{table}

\noindent \textbf{Demand-aware peer selection.} 
In one of the early works on demand-aware peer selection, the authors of~\cite{aggarwal2007can} focus on accounting for the underlying Internet topology, and thus leveraging the cooperation with internet service providers.
Among the most recent related works, we note~\cite{WangKZGJM0K23}, which introduces a demand-aware peer-selection algorithm for training deep neural networks. In this work, we refer to the algorithm in~\cite{WangKZGJM0K23} as the \emph{Permutations algorithm}. Intuitively, this algorithm consists of $\log n$ overlapping rings of peers on top of each other, and depends on coin-change routing. This routing is not in line with the traditional XOR-based routing, hence can not be easily integrated with existing protocols. Furthermore, as this work only focuses on overlapping rings, it does not fully utilize many other possible peer-selection scenarios.
Lastly, we want to point out that in the context of blockchain systems, works such as~\cite{BabelB22,MaoDVKS20} optimize the peer-selection procedure to minimize latency between the broadcast of a transaction and its confirmation. Compared to these works, our objective of adjusting to any network traffic demand is more general. 

In the empirical evaluation, we compare our work to Chord and to the Permutations algorithms. A comparison of the theoretical guarantees of these algorithms is presented in Table~\ref{table:DHT}.

\section{Model}
In this section, we discuss the underlying model used in the design of our algorithm and in the empirical analysis.

The peer-to-peer network considered in this paper is an overlay topology operating on top of an underlying network.
We consider an overlay network topology $N$ consisting of $n$ peers/nodes\footnote{From now on, we use nodes and peers interchangeably.} (e.g., computers) with identifiers $0, 1, \dots, n-1$. Without loss of generality, we assume that $n$ is a power of two. Note that the algorithms presented in this paper assign nodes to buckets of different sizes. If $n$ was not a power of two, the algorithms would simply assign the additional nodes to the largest bucket. Further, we assume that the nodes of the network communicate with each other by sending messages over the neighboring links and forwarding other nodes' messages. 

\noindent \textbf{Overlay network.} We consider an overlay network that consists of two parts. 
First, to ensure the connectivity of the underlying network, we consider the existence of an overlay network structure where the $n$ nodes are already connected to each other such that they form a ring. This is a common assumption in the literature~\cite{StoicaMLKKDB03}.
Then, to achieve an efficient peer-to-peer network, we allow nodes to locally augment the network by adding up to $\log n$ additional links to their peers. We assume that the links of these two parts have the same properties: all having equal capacity and delay between their two endpoints.
We denote the distance between nodes $i$ and $j$ by $\dis_{i,j}$. The exact value of the distance depends on the underlying routing strategy, detailed below.

\noindent \textbf{Routing mechanisms.} 
We define routing on the overlay network (including the ring and the augmented links) and discuss two possible routing mechanisms.
The first considered routing strategy uses the \emph{shortest paths}. This is a routing strategy that always uses the smallest number of edges for communication. When computing the shortest path distance, we assume that all nodes are aware of the whole network. We only consider this type of routing as a benchmark for other routing strategies, since assuming the knowledge of the whole network is unrealistic.

The second routing strategy is \emph{local routing}. Here, we assume that each node has a local view of the network. The nodes forward messages based on the messages they receive and their own connections to neighbors. As many local routing strategies depend on the peer-selection algorithm, we will specify these strategies later in the paper. Observe that local routing, unlike shortest path routing, only require nodes to keep track of their neighbors. 

\noindent \textbf{Demand matrix.} To be able to design demand-aware peer-selection algorithms, we assume that the communication demand of the network is fully described by a demand matrix $D \in \mathbb{R}^{n \times n}$. Each entry $i$ and $j$ of the demand matrix shows the amount (or the percentage of) traffic between nodes $i$ and $j$. Observe that the elements on the diagonal of the demand matrix always have zero values, as the nodes do not communicate with themselves. Moreover, since we consider directed networks, the demand matrix is not required to be symmetric. 

\noindent \textbf{Cost function.} The objective of our model is to minimize the overall communication cost. This cost depends on the communication matrix and on the routing strategy in the augmented network. Formally, the cost function $C_N$ for a network $N$ is defined as
$$C_N = \sum \limits_{\forall i, j} \dis_{ij} \times D_{ij}$$
where $\dis_{ij}$ is the distance between the nodes $i$ and $j$ under the applied routing strategy.

\section{Peer-Selection Algorithms}

\begin{figure}
	\centering
	\includegraphics[scale=0.7]{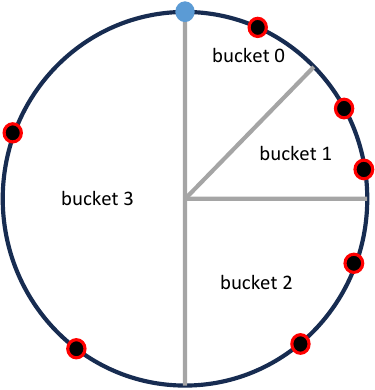}
	\caption{This figure shows the layout of the buckets on a cycle from the blue node's perspective.  We assume that the blue node has the key $0000$. The buckets from the farthest to the closest relative to the source node (the blue one) are correspondingly indexed from 0 up to $\log n - 1$. They are made up of all nodes with prefixes $0001$, $001$, $01$, and $1$ respectively.}
	\label{fig:buckets}
\end{figure}

We now describe our demand-aware peer-selection algorithms. The overarching part of our algorithms is inspired by the Kademlia protocol~\cite{MaymounkovM02}. In the routing tables of Kademlia, nodes are organized in a binary tree-like structure. For our algorithms, we instead use the cyclic structure.

From the perspective of a single source node, the entire key-space is first divided into $\log n$ buckets, as shown in an example in Figure~\ref{fig:buckets}.
Bucket $0$ only contains the successor node, while the largest and farthest bucket contains half of the nodes inside the network. As the nodes in a bucket always have keys with a specific length of common prefix, a bucket can be simply represented by a key range with its start point and end point. Algorithm~\ref{alg: bsb} presents this overarching part as pseudocode. 
The next step is to select a peer in each bucket.
We have designed two demand-aware methods for peer selection in a bucket. A summary of these two node selection algorithms is described below and is presented as pseudocode in Algorithm~\ref{alg: LocatNode}. 

\begin{algorithm}[t]
	\caption{\bsb algorithm}
	\label{alg: bsb}
	\begin{algorithmic}[1]
		\STATE $degree \gets \log n$
		\FOR{$i \gets 1 \; to \; n$}

		\FOR{$j \gets 1 \; to \; degree$}
		\STATE $startpoint \gets (i \oplus (1 \ll j)) \gg j \ll j$
		\STATE $endpoint \gets startpoint + (1 \ll j) - 1$
		\STATE $demand \gets D[i][startpoint:endpoint]$
		\STATE $peer \gets LocateNode(n, i, demand)$

		\STATE $peer \gets (indexOfMax(demand) + i) \% n$
		\STATE $G[i][peer] \gets 1$
		\ENDFOR
		\ENDFOR
	\end{algorithmic}
\end{algorithm}

\begin{algorithm}
	\caption{LocateNode$(n,i,demand)$}
	\label{alg: LocatNode}
	\begin{algorithmic}[1]
        \IF{\halfsplit}
		\STATE $total\_demand \gets 0$
		\FOR{$k \gets 1 \; to \; n$}
		\STATE $total\_demand \mathrel{+}= demand[k]$
		\ENDFOR
		\STATE $half\_demand \gets \frac{total\_demand}{2}$
		\STATE $cumulative\_demand \gets 0$
		\FOR{$k \gets 1 \; to \; n$}
		\STATE $cumulative\_demand \mathrel{+}= demand[k]$
		\IF{$cumulative\_demand \geq half\_demand$}
		\STATE $node = (i + k) \mod n$
		\ENDIF
		\ENDFOR
        \ENDIF
        \IF{\maxdemand}
        \STATE $max\_demand \gets -1$
        \FOR{$k \gets 1 \; to \; n$}
        \IF{$demand[k] > max\_demand$}
            \STATE $max\_demand = demand[k]$
            \STATE $node = (i + k) \mod n$
        \ENDIF
        \ENDFOR
        \ENDIF
        \RETURN $node$
	\end{algorithmic}
\end{algorithm}

\subsection{Half-split}
Given the local communication demand from the source node $s$, from which the corresponding bucket-level demand can be extracted, node $s$ chooses the peer in each bucket such that it splits the bucket-level demand approximately evenly. Note that bucket $0$ only contains the successor node of the source node, so node $s$ will directly add its successor in bucket $0$ to the routing table. In other buckets, node $s$ determines the node along the cycle that has approximately $50\%$ of the total bucket-level demand on both sides, i.e., splits the bucket-level demand in half. We call this node the \textit{mid-node}, and a link to this node will be added to the source's routing table. 

\subsection{Max-demand}
Another demand-aware method for peer selection is to directly establish links to the node that has the highest inbound demand from the source node. Intuitively, this \maxdemand method could help shorten the routing path length between node pairs that communicate frequently, since it could at least reduce the communication cost from the source node to peers within its routing table. Note, however, that finding the peer with the highest bucket-level demand is more time-consuming than the previous method.

\subsection{XOR-based greedy routing}
Since the peers are chosen in a demand-aware manner, the global routing information is not available to the nodes. Therefore, a node can only determine the next hop based on the destination key. In other words, the message will be forwarded to the node whose key has the longest common prefix with the destination key in the current node's routing table. 
Unlike the shortest path routing in a peer-to-peer network, the greedy routing does not require individual nodes to know all links. At each step of forwarding a message, the node only considers the destination key to determine the next hop.

The pseudo-code for XOR-based greedy routing is presented in Algorithm \ref{Algo Prefix-based Routing}. In this algorithm, each node at each step checks the peers in its routing table and chooses the peer with the longest common prefix with the destination key as the next hop, until the message arrives at the destination. Consequently, the length of the XOR-based routing path is upper bounded by $\log n$, as the key is represented by a binary value of $\log n$ bits.

\begin{algorithm}[tb]
	\renewcommand{\algorithmicrequire}{\textbf{Input}}
	\renewcommand{\algorithmicensure}{\textbf{Output}}
	\caption{XOR-based geedy routing}
	\label{Algo Prefix-based Routing}
	\begin{algorithmic}[1]
		\REQUIRE $n:$ Number of nodes in the network $(n=2^m, m \in \mathbf{N}) $
		\REQUIRE $G:$ Final topology with directed edges $(G\in \{0, 1\}^{n \times n})$
		\ENSURE $R: $ XOR-based routing path lengths in a matrix$(R\in \mathbf{R}^{n \times n})$
		\STATE \emph{\textcolor{blue}{$\triangleright$ Compute the path length from each node-pair and add the corresponding routing cost to total cost}}
		\FOR{$i \gets 1 \; to \; n$}
		\FOR{$j \gets 1 \; to \; n$}
		\STATE $curr \gets i$
		\STATE $next\_hop  \gets curr$
		\STATE $max\_common\_prefix\_len \gets 0$
		\STATE $path\_length \gets 0$
		\STATE \emph{\textcolor{blue}{$\triangleright$ Stop when the destination node $j$ is arrived}}
		\WHILE{$curr \neq j$}
		\FOR{$k \gets 1 \; to \; n$}
		\IF{$G[curr][k] = 1$}
		\STATE $common\_prefix\_len \gets log_2n - bit\_length(k \oplus j)$
		\IF{$common\_prefix\_len > max\_common\_prefix\_len$}
		\STATE $max\_common\_prefix\_len \gets common\_prefix\_len$
		\STATE $next\_hop \gets k$
		\ENDIF
		\ENDIF
		\ENDFOR
		\STATE $curr \gets next\_hop$
		\STATE $path\_length \mathrel{+}= 1$
		\ENDWHILE
		\STATE \emph{\textcolor{blue}{$\triangleright$ Add the routing cost from node $i$ to node $j$}}
		\STATE $R[i][j] \gets path\_length$
		\ENDFOR
		\ENDFOR
		\STATE \RETURN $R$
	\end{algorithmic}
\end{algorithm}

\section{Experimental Evaluation}

The main goal of the evaluation section is to answer the following questions:

\begin{itemize} 
    \item \textbf{Question 1:} What is the running time of the presented peer-selection algorithms?
    \item \textbf{Question 2:} Under which parameters do the algorithms perform best? 
    \item \textbf{Question 3:} How do our algorithms perform in a wide range of real-world datasets?
\end{itemize}

\subsection{Datasets}
In our work, we consider two variants of datasets, a synthetic dataset that was created based on Zipf distribution, and also the realistic dataset from~\cite{AvinGGS20}.

\noindent \textbf{Zipf distribution.} We generated 16 synthetic network communication data samples with different $\alpha$-values, following the rules of the random Zipf distribution \cite{Zipf2}. $\alpha$ is a hyperparameter in generating sequences based on Zipf distribution, which must be greater than $1$. The probability density of every random variable becomes smaller as the $\alpha$-value increases. That is, the skewness of the distribution becomes weaker with larger $\alpha$-values.

\noindent \textbf{Real-world dataset.} 
We conducted our simulations using two of the three datasets available: cluster A and cluster C. They are collected from two different applications, namely Database and Hadoop applications. Every split data chunk includes network packets in a 20-minute time interval, where the time intervals of different data chunks do not overlap with each other.

Besides the Facebook main datasets, we evaluated our work on datasets of smaller sizes from various applications, like Microsoft, pFabric, and ProjecToR.

\begin{figure*}[t]
        \captionsetup[subfigure]{justification=centering}
	\centering
	\begin{subfigure}[b]{.47\textwidth}
		\centering
		\includegraphics[width=\linewidth]{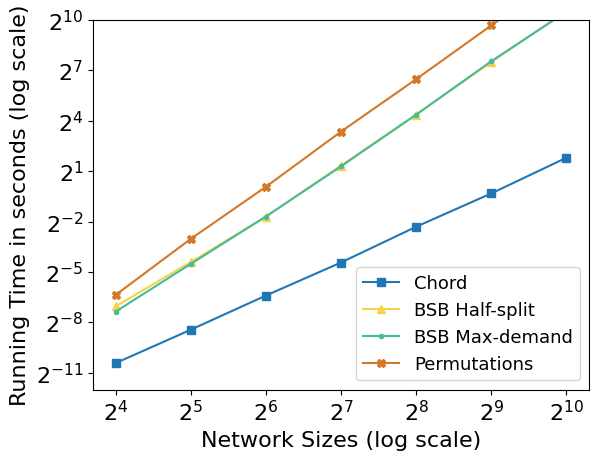}
		\caption{Running time}
        \label{fig: time}
	\end{subfigure}
 	\begin{subfigure}[b]{.52\textwidth}
		\centering
		\includegraphics[width=\linewidth, clip]{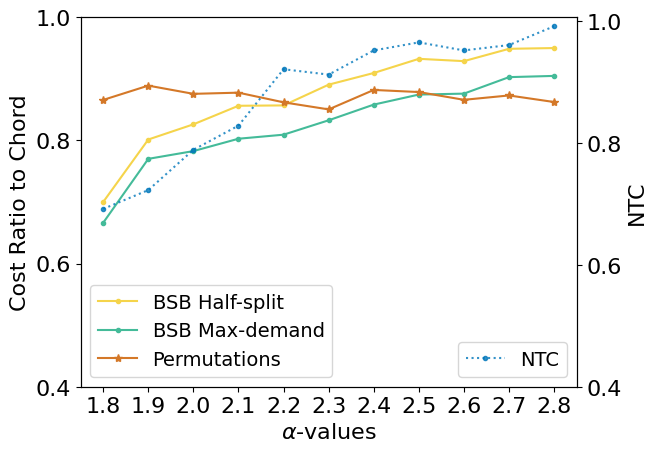}
		\caption{Cost compared to Chord}
        \label{fig: non-temporal}
	\end{subfigure}
	\caption{Figure~\ref{fig: time} demonstrates the running time of the peer-selection algorithms that we focus on. Figure~\ref{fig: non-temporal} shows the correlation between the communication costs calculated with different peer-selection algorithms and the non-temporal complexity (NTC) of the network traffic data.}
    \label{fig: results}
\end{figure*}

\subsection{State-of-the-art algorithms}
In the following, we provide a short overview of how we implemented two peer-selection algorithms from the literature, Chord~\cite{StoicaMLKKDB03} and the Permutations algorithm~\cite{WangKZGJM0K23}, and also how the coin-change routing algorithm works.

\noindent \textbf{Chord.} 
Chord employs a cyclic topology of nodes. 
In this case, nodes simply select peers using the bitwise XOR operator. That is, the $k$-th peer of node with ID $i$ is $(i \oplus 2^k) $. Moreover, the routing between any pair of nodes on a cyclic network of size $n$ is guaranteed within $\log n$ steps.

\noindent \textbf{Permutations algorithm.} 
This algorithm is similar to the standard Chord protocol in the sense that the peers are selected based on a group of predefined unidirectional distances to the source node. The difference is that while the distances in the Chord protocol are powers of two, those in the permutation algorithm are calculated based on the given demand matrix. The permutations are selected from the integers in $[1, n-1]$.
Then the GCD-filter (greatest common divisor filter) is applied to reduce the number of candidate permutations. The final permutations that define the augmented communication links are chosen one by one from the candidate list. The network is then updated with additional edges after each selection. The GCD-filter ensures the connectivity of the network during the process of permutation selection, which is essential for the calculation of communication costs.

\noindent \textbf{Coin change algorithm.} The goal of the coin change algorithm is to find a way to deliver a target amount of money using the minimum number of coins. The available coins refer to the selected permutations, and the target amount of money refers to the unidirectional distance from the source node to the destination node. The coin change algorithm computes the minimum number of permutations needed to have all possible distances ranging from $1$ to $n - 1$. It correspondingly stores the combinations of permutations that make up the routing paths associated with valid distances. Dynamic programming is used to reduce the computational cost of repeatedly calculating the routing paths for each source-destination pair. Once the calculation is finished, routing becomes quite fast by looking up the associated linear combination of permutations based on distances in a list of $n-1$ elements.

\subsection{Results}
To simulate our system, we have used python~$3.10$ with networkx~\cite{NetworkX} and Matplotlib~\cite{Matplotlib} libraries. Our programs were running on $16$ AMD Opteron™ processors with a clock frequency of 2.0GHz, and $15$ GB of RAM.\footnote{The source code of this paper can be found in \url{https://github.com/inet-tub/BSB}}

\noindent \textbf{Answer 1. Running time.}  As shown in Figure~\ref{fig: time}, our \bsb algorithms have an advantage over the Permutations algorithm due to faster computational speed, which makes them stand out in some specific domains.

\noindent \textbf{Answer 2. Non-temporal complexity.}
In order to compute the non-temporal complexity of the datasets, we prepared three data files per network and compressed the files, respectively, following the methodology of~\cite{AvinGGS20}. Firstly, the original file stores the source-destination pairs associated with the packets transmitted across the network in their original sequence. On the basis of the original file, another file, the shuffled file, is generated by randomly shuffling the rows in the original file. At this point, the temporal structure of the network traffic is eliminated. The third file consists of the same number of rows as the original and shuffled file, while the source-destination pairs are randomly generated from the node set using the uniform distribution. During this step, the non-temporal structure is then removed. We name it the random file.
Therefore, the non-temporal complexity is measured as the ratio of the size of the compressed shuffled file to that of the compressed random file.
Figure~\ref{fig: non-temporal} depicts a strong correlation between the cost ratios of \bsb algorithms and the non-temporal complexity of communication demand, which indicates that our \bsb algorithms are targeting cases of low non-temporal complexity.

\begin{figure}[t]
\centering
	\includegraphics[width=0.5\linewidth, clip]{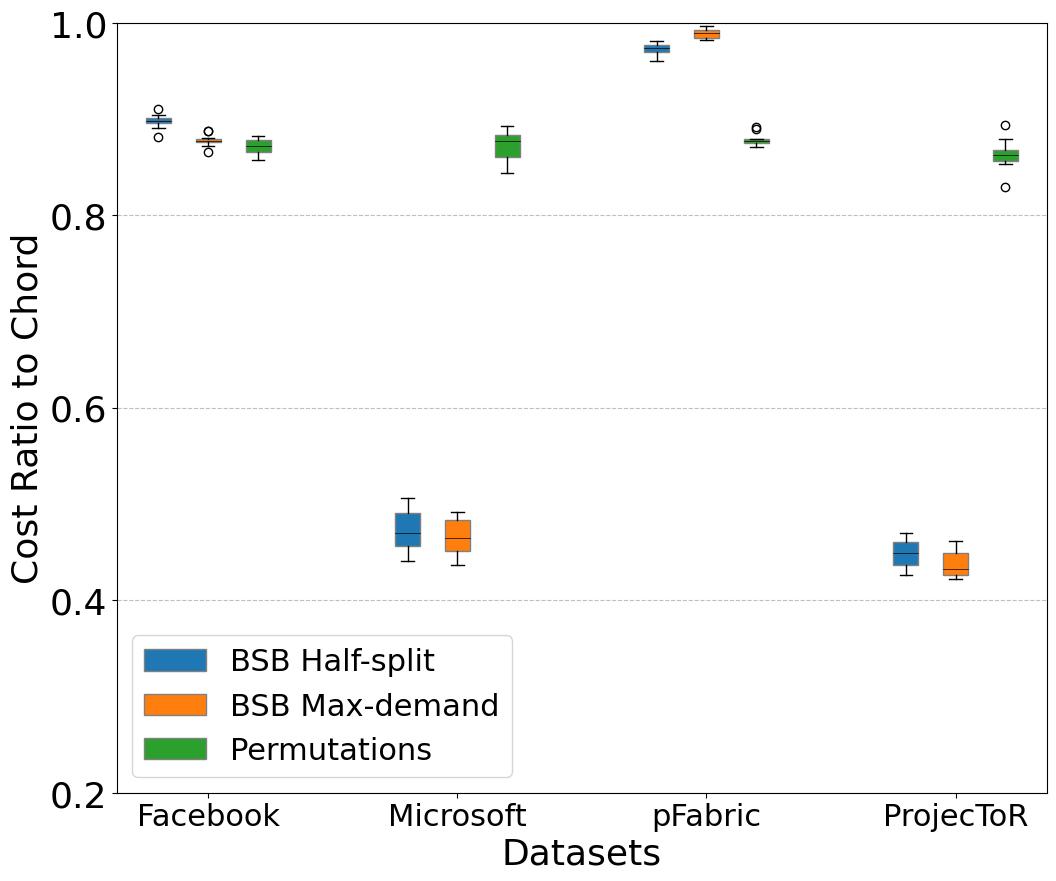}
	\caption{
    Communication cost ratio of \bsb and Permutations algorithms compared to Chord, considering various small networks (with 64 nodes in each).}
    \label{fig:Figure Boxplot Small}
\end{figure}

\noindent \textbf{Answer 3.1. Small real-world dataset.}
Figure~\ref{fig:Figure Boxplot Small} exhibits the simulation results on smaller datasets. We can see the four groups on the x-axis, namely Facebook, Microsoft, pFabric, and ProjecToR. The original datasets are all based on a $100$-node peer-to-peer network. To simplify the process of XOR-based routing, we reduced the number of nodes to $64$ in each group.
We performed the filtering as follows: each of the original $100$ nodes is randomly assigned a node identifier that ranges from $0$ to $99$. Then the nodes with identifiers from $0$ to $63$ were selected. After that, we walked through the whole dataset to remove the rows (i.e., network packets) where either the source node or the destination node is out of the filtered scope.
For each application, we performed $10$ random ID assignments to avoid the situation where a single result is not sufficiently representative for our study. 

Similarly, the performance of the Permutations algorithm does not vary much on different datasets, with a cost ratio of $0.86$ on average. Nevertheless, the simulations of our \bsb algorithms show completely different results in different types of applications. On Microsoft and ProjecToR networks, our \bsb algorithms clearly outperform the Permutations algorithm with an average cost reduction of $55\%$ (compared to Chord). On the other hand, they have only a slight advantage over Chord on Facebook and pFabric networks, underperforming compared to the Permutations algorithm. This phenomenon is again related to our hypothesis that the non-temporal complexity of the network flow could affect the efficiency of our \bsb peer selection. In other words, we infer that the frequency distribution of network communications in Microsoft and ProjecToR datasets seems to be more skewed than that in Facebook and pFabric datasets.

\begin{figure}[t]
        \captionsetup[subfigure]{justification=centering}
	\centering
 	\begin{subfigure}[b]{.49\textwidth}
		\centering
\includegraphics[width=\linewidth]{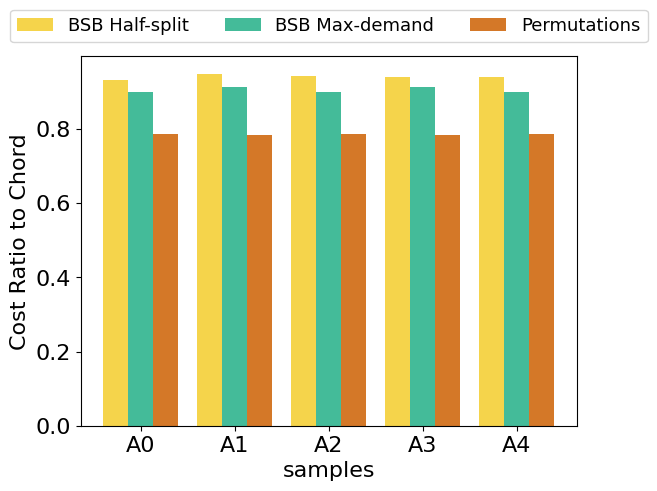}
		\caption{Facebook cluster A}
        \label{fig: facebook A}
	\end{subfigure}
  	\begin{subfigure}[b]{.49\textwidth}
		\centering
\includegraphics[width=\linewidth]{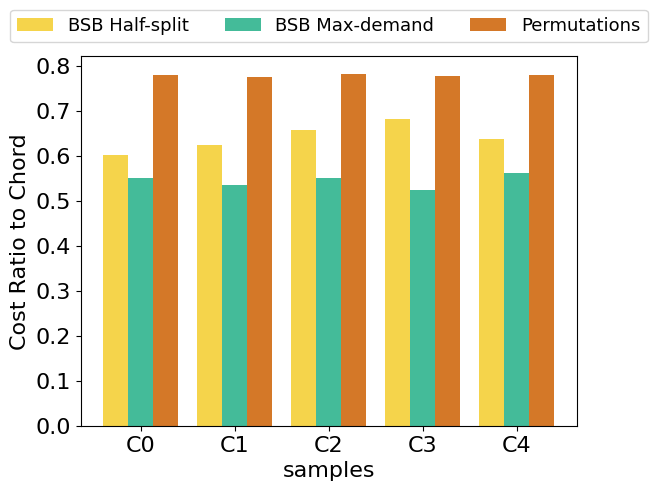}
		\caption{Facebook cluster C}
        \label{fig: facebook C}
	\end{subfigure}
    \caption{This figure shows the communication cost ratio of \bsb and Permutations algorithms compared to the Chord algorithm, which are simulated on 4096-node networks generated from Facebook data samples. Figure~\ref{fig: facebook A} shows results for the database cluster (A). Figure~\ref{fig: facebook C} shows results for Hadoop cluster (C).}
	\label{fig:Facebook}
\end{figure}

\noindent \textbf{Answer 3.2. Real-world dataset.} Each full dataset is partitioned into data chunks based on the timestamps of the network packets, where each data chunk includes the network traffic data within a non-overlapping 20-minute time interval. Although the network size is reduced to 4096 in each data chunk using a key-filter, the cost calculation is still a time-consuming task, especially for the Permutation peer-selection algorithm. Therefore, we conducted simulations only on five data chunks per Facebook cluster due to the time limit.

Although our \bsb algorithms and the Permutations algorithm are both based on the real-world network traffic, their peer-selection logics are quite different. As we discussed in the previous section, the peer-selection method is highly dependent on the routing mechanism.
What we can now observe from the two plots is that the Permutations algorithm shows a stable performance on both datasets. Its cost ratio stays around $0.78$, which indicates that it can reduce the communication cost by $22\%$ compared to the Chord algorithm. 

However, our \bsb algorithms do not present results similar to the Permutations algorithm. They perform much worse than the Permutations algorithm on the Database cluster, whereas they perform much better on the Hadoop cluster. The \bsb \halfsplit algorithm exhibits a cost ratio of approximately $93\%$ on the former dataset and $60\%$ on the latter, where the \bsb \maxdemand algorithm shows lower cost ratios of around $89\%$ and $55\%$ respectively.
A reason for the unstable performance of our \bsb algorithms could be the different frequency distributions of communication demand in various applications.

\subsection{Discussion}

\noindent \textbf{\bsb algorithms versus the Permutations algorithm.}
Intuitively, if there is a high communication demand between a source-destination pair, the easiest way to ensure low cost is to connect the source node to the destination nodes directly.
Since the \bsb algorithms use the \quoteMe{local view}, simply connecting such source-destination pairs would be possible. Especially in the \maxdemand version, the direct channels from the source node to its most frequent contacts in each bucket could be very helpful for reducing the total overhead. However, selecting peers using the \quoteMe{global view} seems to be more complicated, as nodes have to ensure that behaviors benefit themselves are also compatible with the collective interests. Nodes in the Permutations algorithm share the same group of permutations, which makes it restrictive to establish a connection for a single source-destination pair, unless those pairs have significantly higher communication demand among other pairs of nodes.

\noindent \textbf{Comparing two \bsb algorithms.}
The reason why \maxdemand performs better than \halfsplit in most of the cases seems to be straightforward: \maxdemand tends to build a direct link from a source node to the node that it has highest demand to. Recall that the total communication cost can be viewed as the sum of weighted path lengths. Hence, by shortening the path lengths between these pairs of nodes that communicate frequently, the communication cost is expected to be heavily reduced. Although \maxdemand is more beneficial for improving  the efficiency of P2P network communication, it has longer runtime than \halfsplit does on selecting peers at the bucket level.

\section{Conclusion}
In this work, we took initial steps toward a demand-aware peer-selection algorithm, that supports XOR-based routing, and aims to minimize the shortest path, weighted by the communication demand between peers. 
In particular, we introduced a class of algorithms under the general name of \bsb, and gave a comparative performance analysis of this class considering a wide range of real-world and synthetic inputs. Our results indicated that when the communication demand is skewed, \bsb algorithms achieved significantly reduced communication cost compared to state-of-the-art algorithms, by up to $43\%$.
In the future, we aim to study extended variants of \bsb algorithms, in particular, algorithms that incorporate randomization. We also aim to extend our evaluation to other applications where peer selection is relevant, for example, in blockchain networks.

\paragraph*{Funding.} This project has received funding from the European Research Council
(ERC) under grant agreement No. 864228 (AdjustNet), 2020-2025.

\bibliography{p2p.bib}

\end{document}